\documentclass[manuscript]{aastex}

\slugcomment{Not to appear in Nonlearned J., 45.}

\shorttitle{Sympathetic solar filament eruptions}
\shortauthors{Rui Wang, et al.}

\begin{document}

\title{Sympathetic Solar Filament Eruptions on 2015 March 15}

\author{Rui Wang\altaffilmark{1}, Ying D. Liu\altaffilmark{1,2}, Ivan Zimovets\altaffilmark{1,3,4}, Huidong Hu\altaffilmark{1,2}, Xinghua Dai\altaffilmark{5}, and Zhongwei Yang\altaffilmark{1}}


\altaffiltext{1}{State Key Laboratory of Space Weather, National Space Science Center, Chinese Academy of Sciences, Beijing 100190, China; liuxying@spaceweather.ac.cn}
\altaffiltext{2}{University of Chinese Academy of Sciences, Beijing 100049, China}
\altaffiltext{3}{Space Research Institute of Russian Academy of Sciences, Profsoyuznaya St. 84/32, Moscow 117997, Russia}
\altaffiltext{4}{International Space Science Institute, Beijing 100190, China}
\altaffiltext{5}{Key Laboratory of Solar Activity, National Astronomical Observatories, Chinese Academy of Sciences, Beijing 100012, China}


\begin{abstract}
The 2015 March 15 coronal mass ejection as one of the two that together drove the largest geomagnetic storm of solar cycle 24 so far was associated with sympathetic filament eruptions. We investigate the relations between the different filaments involved in the eruption. A surge-like small-scale filament motion is confirmed as the trigger that initiated the erupting filament with multi-wavelength observations and using a forced magnetic field extrapolation method. When the erupting filament moved to an open magnetic field region, it experienced an obvious acceleration process and was accompanied by a C-class flare and the rise of another larger filament that eventually failed to erupt. We measure the decay index of the background magnetic field, which presents a critical height of $118$ Mm. Combining with a potential field source surface extrapolation method, we analyze the distributions of the large-scale magnetic field, which indicates that the open magnetic field region may provide a favorable condition for F2 rapid acceleration and have some relation with the largest solar storm. The comparison between the successful and failed filament eruptions suggests that the confining magnetic field plays an important role in the preconditions for an eruption.
\end{abstract}




\section{INTRODUCTION}
Solar eruptions are often observed as filament eruptions, flares, and coronal mass ejections (CMEs). These dynamic phenomena are usually considered to be connected with each other, and they may be different manifestations of the same magnetic energy release process in the corona. Solar filaments are relatively cool and dense material suspended in the corona. They are dark and uneven strips observed against the solar disk and called a prominence observed on the limb. CMEs have strong relations with filament/prominence eruptions. Part of the erupting filament becomes the bright core of the CME, with the remaining part falling down to the solar surface. The occurrence of a CME generally relates to the disruption of magnetic equilibrium. The imbalance between the downward magnetic tension force and the upward magnetic pressure force would cause the CME progenitor (generally a strongly twisted or sheared magnetic structure, e.g., the flux-rope structure in a filament channel) to be unstable. There are various ways the imbalance is caused, e.g., shearing motion \citep{1990Aly}, emerging flux \citep{1995Feynman}, torus \citep{2006Kliem} or kink \citep{2004Torok} MHD instability, breakout reconnections \citep{1999Antiochos}, or catastrophic loss of equilibrium \citep{1991Forbes}. These mechanisms would make the CME progenitor from an equilibrium state into a metastable state \citep{2001Sturrock} or a state very close to the loss of equilibrium \citep{2000Forbes}. For the latter case, the CME eruption can be triggered by small perturbations, e.g., magnetic twist, but for the former case, more powerful triggers are necessary for the eruption, such as flares or filament eruptions as the triggers.

The interactions between two filaments can cause their eruptions and lead to a flare or CME \citep{2007Su,2010Kumar,2014Jiang,2015Zhu}. The interactions often indicate that magnetic reconnection occurs between two flux-rope systems. \citet{2001Linton} proposed four types of flux tube interactions: bounce, merge, slingshot, and tunnel. \citet{2015Zhu} reported a convergence and direct interactions of a erupting lower filament with the other descending upper one. A complicated eruptive structure was formed by the merging of the two filaments and finally erupted away.

On the other hand, the appearance of one eruption in one active region (AR) can trigger the appearance of the other eruption in the other region. The triggering process can achieve through the destabilization by the large-scale convective motions \citep{1993Bumba} or perturbations along the interconnecting field lines \citep{2008Jiang}. People call this kind of event a sympathetic event. Compared with the filament--filament interactions, the sympathetic filament eruptions generally occur within a relatively shorter period of time in two different physically related regions \citep{1993Bumba,2001Wang,2011Schrijver,2011Torok,2012Titov,2016Joshi}. Except for the triggering mechanisms of the sympathetic eruptions mentioned above, \citet{2011Torok} studied the sympathetic eruptions of three filament systems and presented a three-dimensional MHD simulation, which suggested that the overlying flux removal of the flux ropes led by a filament eruption resulted in the consecutive filament eruptions.

On 2015 March 15, a small-scale filament eruption presenting surge-like characteristics in the AR 12297 failed to erupt, but induced another suspended filament to erupt. The erupting filament evolved into a fast CME, which together with another CME $12$ hr earlier on March 14 drove the $largest$ geomagnetic storm of solar cycle 24 so far with a $D_{st}$ minima of $-223$ nT \citep{2015Liu}. The source region itself is interesting, and we will focus on the investigation of the source region in this Letter and see what happened on the Sun. First, we try to confirm that this event is a sympathetic filament eruption with multi-wavelength observations and using magnetic field extrapolation methods. Second, we investigate the relation between the magnetic topology and acceleration processes of the erupting filament. The results will also help us better understand the trigger and acceleration mechanisms of CMEs.

\section{OBSERVATIONS AND DATA ANALYSIS}\label{observations}

On 2015 March 14 and 15 there occurred two CMEs. The first CME is a partial halo CME associated with a C2.6 flare and first appeared in LASCO C2 around 13:30 UT on March 14. The second CME is a halo CME associated with a C9.1 flare and first appeared in LASCO C2 at 01:48 UT on March 15. Our focus is the second CME eruption with a maximum speed of about 1100 km s$^{-1}$ \citep{2015Liu}, which was involved with sympathetic filament eruptions. Figure 1a shows that there existed at least four filaments around AR 12297. Filament 1 (F1) erupted appearing like a surge eruption. Filament 2 (F2) was located nearby the AR, one of whose ends was rooted in the AR; the other end is outside the AR, which can be observed during the brightening around the end (see the black circles of Figure 1b). F2 was the erupting filament that was finally triggered into the March 15 CME. Filament 3 (F3) was a dark S-shaped filament channel, one of whose ends was rooted near the AR. F3 rose up during the eruption and exhibited large-scale twisted loop structures, but this is a failed eruption. Filament 4 (F4) was a quiet filament channel that had already erupted during the first CME eruption on March 14. It remained almost static before and after the F1 eruption. Figure 1c shows the CMEs seen in the LASCO C3 view. The March 15 CME had a major component heading west. It interacted with the first largely southward propagating CME, which contributed to the occurrence of the intense geomagnetic storm of March 17 \citep{2015Liu}.

\subsection{The Surge-like Filament Eruption}
Small-scale filament eruptions are often accompanied with transient phenomena, e.g., solar flares, H$_\alpha$ surges, or X-ray jets. Small-scale filament eruptions are easily confused with the H$_\alpha$ surge eruptions morphologically. H$_\alpha$ surges are straight or slightly curved ejections of chromosphere matter into coronal heights, and the ejected matter often returns back along almost the same path as in the rising phase \citep{1973Roy,1977Bruzek}. Figure 2a shows the line of sight (LOS) magnetogram of the AR from the Helioseismic and Magnetic Imager (HMI; \citealp{2012schou}) on board the {\it Solar Dynamics Observatory} ({\it SDO}; \citealp{2012pesnell}). The region in the yellow box locates F1 before the eruption. The distribution of the photospheric magnetic field in the box exhibited a single negative-polarity region surrounded by positive polarities and formed a circular magnetic polarity inversion line (PIL). Therefore, we call this region a circular magnetic field region (CR). This kind of distribution of the magnetic field is a favorable condition for the 3-dimensional reconnection discussed by \citet{2009Masson} and \citet{2010Pariat} and is also likely to form a jet \citep{1995Yokoyama,2013Archontis}. Figure 2b displays the flare ribbons in the 1600 \AA~image of the Atmospheric Imaging Assembly (AIA; \citealp{2012lemen}) on {\it SDO}, which were led by the C2.4 flare beginning at 00:35 UT. The flare ribbons consist of a center ribbon and a circular ribbon, which are consistent with the distribution of the photospheric magnetic field. We also used H$_\alpha$ images from the Global Oscillation Network Group (GONG) program. The H$_\alpha$ images in Figure 2c and d clearly show F1, F2, and F3 before and after the eruption, respectively. The small-scale filament F1 existed in Figure 2c but disappeared in Figure 2d. Meanwhile, a surge-like eruption appeared that originated from the vicinity of F1. Figure 2e exhibits F1 as a small S-shaped dark feature lying on the PIL of the CR at AIA 304 \AA~. Figure 2f gives the reconstructed magnetic fields along the same region. As a small-scale filament, F1 should be very close to the non-force-free photosphere, so we used a forced magnetic field extrapolation method \citep{2013Zhu,2016Zhu} that is appropriate for the real situations of the magnetic fields near the surface of the Sun. The extrapolation results exhibit a flux-rope structure held by the overlying field that connected the center negative polarity to the peripheral positive polarity of the CR. The flux-rope structure coincides well with F1 as shown in Figure 2e. We also present the nonlinear force-free field (NLFFF; \citealp{2004wiegelmann,2010wiegelmann}) and potential field extrapolation \citep{1978Seehafer} results for comparisons (see Figure 2g and 2h). The NLFFF extrapolation adopts the same free parameters as Case-E in the work of \citet{2012wiegelmann}. The potential field exhibits a typical fan-spine structure as shown by \citet{2010Pariat}. The NLFFF result presents an outer spine structure with little twisted fields in the region of F1. However, we do not obtain the flux-rope structure by both the methods.

Figure 3a shows the characteristics of the F1 eruption accompanied with the C2.4 flare. The {\it Reuven Ramaty High Energy Solar Spectroscopic Imager} (RHESSI; \citealp{2002Lin}) X-ray (XR) 6 $\--$ 12 keV map (blue contour) overlaid on the AIA 304 \AA~image presents the flare source region that coincides with the CR. The eruption seems to give rise to double-layer ejecta. The upper bright structure looked like a hot X-ray jet and the lower dark one appeared as a surge. \citet{2015Sterling} reported that miniature filament eruptions initiating CMEs can drive jets through the continuous reconnection between the minifilament-carrying erupting fields and the ambient open fields. Based on the extrapolation results, the jets probably erupted along the outer spine shown in Figure 2g and 2h. According to the site of the XR source, the reconnection are likely to occur at the `legs' of F1, and the jet spire drifted away from the X-ray-jet bright point at the edge of the jet's base (see the animation), which are analogous to the process described by \citet{2015Sterling}, rather than the accepted version of the emerging-flux model of jet formation that was associated with null point reconnection \citep{1995Yokoyama,2010Pariat}. The lower surge-like eruption swung southward in Figure 3b. We made a slice along the quasi-perpendicular direction of the motion. Figure 3c is the stack plot of the intensity along the slice. The southward trends of the bright strips can be tracked by the white asterisk lines, which indicate that the failed eruption fell back in the ending phase. The dark ejecta in Figure 3a should consist of the filament-carrying erupting fields but not be an H$_\alpha$ surge since the ejected matter of a surge often returns back along almost the same path as in the rising phase and does not swing in the perpendicular direction. Based on these results, we conclude that the surge-like eruption is the eruption of the small-scale filament F1, not an H$_\alpha$ surge.

\subsection{The Initiation and Acceleration of The Major Filament Eruption}
Figure 4a shows that the erupting F1 destabilized F2 (probably through perturbations along the interconnecting field lines) around 00:40 UT, which then propagated southwestwards in a rolling motion. F2 stopped rolling and was accelerated rapidly around 01:15 UT (Figure 4b). Meanwhile, with the rapid outward propagation of F2, F3 rose up and gradually formed large-scale twisted loops, like a ``dragon head.'' The C9.1 flare accompanied the F2 eruption. We overplotted the RHESSI XR 6 $\--$ 12 keV map (red contour) on the AIA 131 \AA~image, which indicates the source region of the C9.1 flare. As there were no RHESSI XR observations during the acceleration of F2, we selected the closest time in the gradual phase of the flare. The XR source was probably located on the top of the flare arcades (see Figure 4c). We made a slice along the propagation path of F2 (see Figure 1b). The temporal evolution of F2 along this slice is shown as the distance--time plot in Figure 4d. The {\it GOES} soft X-ray (SXR) flux is also overlaid on this plot. Comparing the distance--time measurement with the SXR curve, we found that the onset of F1 eruption coincided with the impulsive phase of the C2.4 flare. It is noteworthy that after the slow rolling and acceleration, F2 showed a further acceleration around 01:15 UT, which is also visible in the velocity and distance curves in Figure 4e.

We examined the global magnetic field configurations around the AR using the potential field source surface (PFSS) model. The whole AR was under a streamer belt and covered by large-scale closed magnetic fields (the white lines in Figure 5a and 5b). Figure 5a shows that F2 stayed under the closed magnetic field before the eruption. However, around 01:15 UT when the acceleration began, F2 reached to the edge of the streamer belt where the open magnetic fields were rooted (the green lines). We put the projected positions (black and blue crosses in Figure 5c) of F2 at two different times together with the open magnetic field region (OR; colored contours).
We think that the acceleration of F2 may result from the change of the background strapping force. To quantitatively describe how strong the background fields were above the starting place of the eruption, a decay index is defined as $n = -d$~log$(B_t)/d~$log($h$) \citep{2006Kliem}, in which $B_t$ is the strength of the strapping field in the transverse direction and computed from the PFSS model and $h$ is the radial height above the photosphere. Figure 5d shows that the critical value $n$ = 1.5 for torus instability onset corresponds to a high altitude (approximately 118 Mm), which indicates that the strapping field is very strong (can refer to the results of \citealt{2008Liu,2012Xu,2015Jing,2015Sun,2015Wangh}  for comparisons). Under such strong background fields, it is not surprising that the S-shaped filament F3 failed to erupt. The acceleration of F2 around 01:15 UT was probably related to the passage of F2 in the region with open magnetic field configurations where F2 was not confined.

\section{CONCLUSIONS}

We have investigated the sympathetic eruptions between different filaments on 2015 March 15, which contributed to the occurrence of the largest geomagnetic storm of solar cycle 24 so far. The small-scale filament F1 lying in a circular magnetic field region erupted, resulting in a C2.4 class flare. The eruption of F1 likely drove a jet through the continuous reconnection between the minifilament-carrying erupting fields and the ambient fields. Although F1 failed to erupt eventually, it made the nearby filament F2 out of equilibrium and induced it into a rolling motion. When F2 propagated into an open magnetic field region, it rapidly accelerated due to the sharp decrease of the constraining force from the overlying magnetic field, which produced the March 15 CME. With its outward propagation, the magnetic field configuration associated with F2 probably experienced a major disruption and was along with a magnetic reconnection process. The C9.1 flare associated with F2 eruption entered into its impulsive phase (see Figure 4d). It was probably a result of the reconnection process developing through the overlying arcade around the flux-rope structure in F2, which was along with the fast expansion of F2 and observed as the XR source shown in Figure 4c. Another filament F3 rose up and formed a twisted loop structure like a ``dragon head,'' but it also failed to erupt. The rising F3 could not overcome the confinement of the overlying closed magnetic fields and failed to erupt. Through the observations and analysis, we summarize the main results as follows:

1. We revealed a sympathetic filament eruption and discussed the acceleration process of the erupting filament. A small-scale filament eruption destabilized a nearby filament. The small-scale filament was reconstructed through a forced magnetic field extrapolation method. We measured the decay index and obtained a critical height of 118 Mm, which reveals that the background magnetic field above the erupting filament is very strong. The open magnetic field region may provide a favorable condition for F2 further acceleration. The erupting filament is likely to propagate along the open magnetic field lines. The nearby coronal hole of the open magnetic field region (see Figure 5a) probably provided high-speed solar wind streams behind the CME, which supports the scenario about the formation of the strong geomagnetic storm on 2015 March 17 proposed by \citet{2015Liu} as a solar surface evidence.

2. Large-scale coronal magnetic fields always play an important role on CME eruption and propagation, such as CME deflection or channeling \citep{2015Mostl,2015Wang}, and sometimes they could decide whether an eruption is a success or failure. There were three filaments in this event. However, only one out of the three erupted. The failure of F1 and F3 eruptions are likely because of the strong overlying confining magnetic fields. The acceleration of F2, which finally erupted, was probably due to the loss of confinement by the overlying magnetic fields. These results provide important clues on the preconditions for a successful filament eruption.

\acknowledgments
The research was supported by the Recruitment Program of Global Experts of China,
NSFC under grant 41374173 and the Specialized Research Fund for State Key Laboratories
of China. We acknowledge the use of data from {\it SDO}, GONG, {\it RHESSI}, {\it GOES}, and {\it SOHO}.

\clearpage



\clearpage
\begin{figure}
\centerline{\includegraphics[width=1\textwidth,clip=]{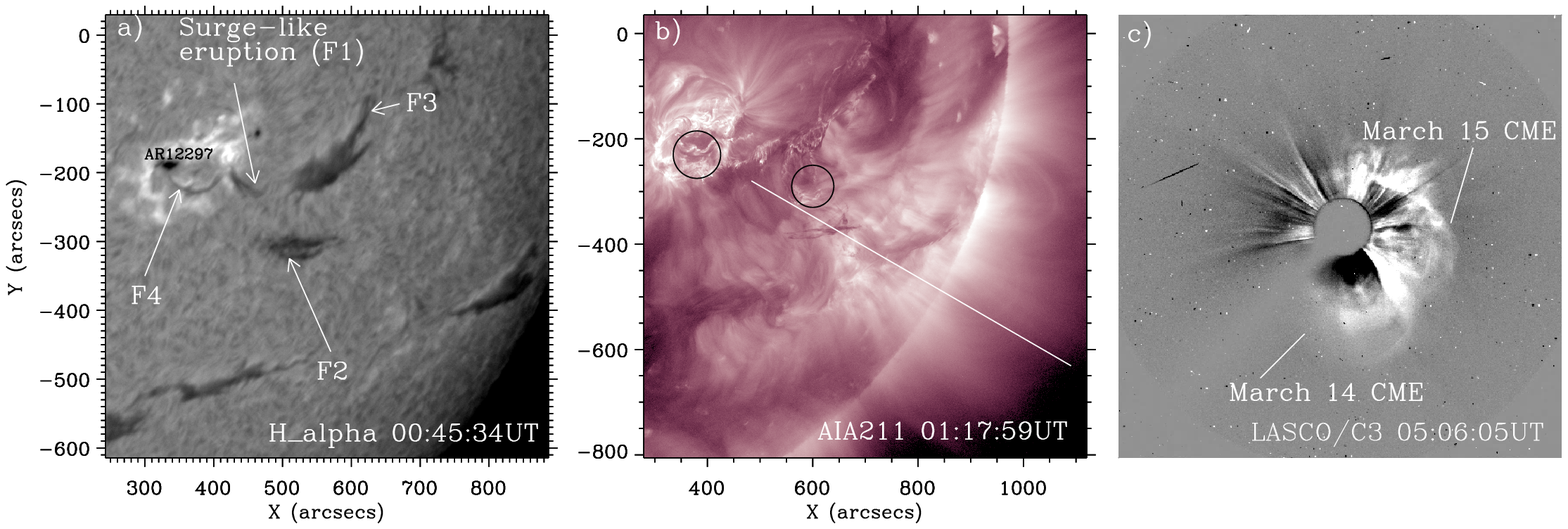}}
\caption{Filaments and the March 15 CME. (a) Different filaments around AR 12297. (b) The white line indicates the slice taken along the F2 progation direction to create the distance--time diagram (see Figure 4d). The black circles indicate the ends of F2. (c) Difference image of the March 15 CME from LASCO C3 on board {\it SOHO}.}
\end{figure}
\clearpage
\begin{figure}
\centerline{\includegraphics[width=0.5\textwidth,clip=]{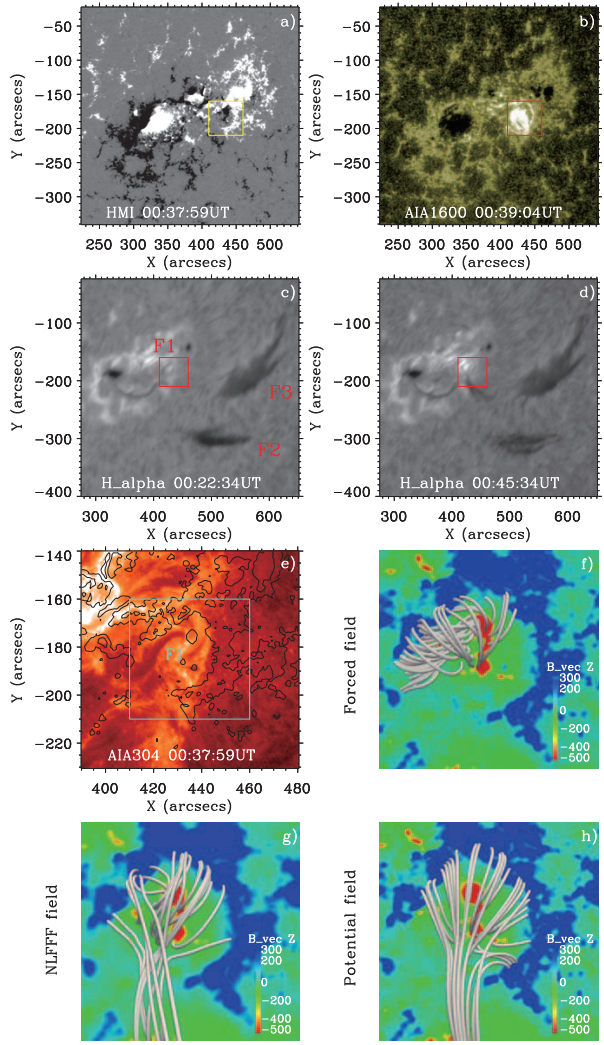}}
\caption{Observations and reconstruction of the small-scale filament F1. (a) {\it SDO}/HMI LOS magnetograms of AR 12297. (b) Circular flare ribbons in the {\it SDO}/AIA 1600 \AA~image. (c) and (d) GONG H$_\alpha$ images before and after the F1 eruption, respectively. (e) AIA 304 \AA~image overlaid by contours of the photospheric LOS magnetic field by HMI. The box in each panel indicates the circular magnetic field region where F1 was located. (f) Forced magnetic field reconstruction around F1 at 00:24 UT. The gray lines indicate the flux-rope structure along F1. The vertical component of the bottom vector magnetic field inside the box region is shown by the color bar. (g) and (h) NLFFF and potential field extrapolation around F1, respectively.}
\end{figure}
\clearpage
\begin{figure}
\centerline{\includegraphics[width=1\textwidth,clip=]{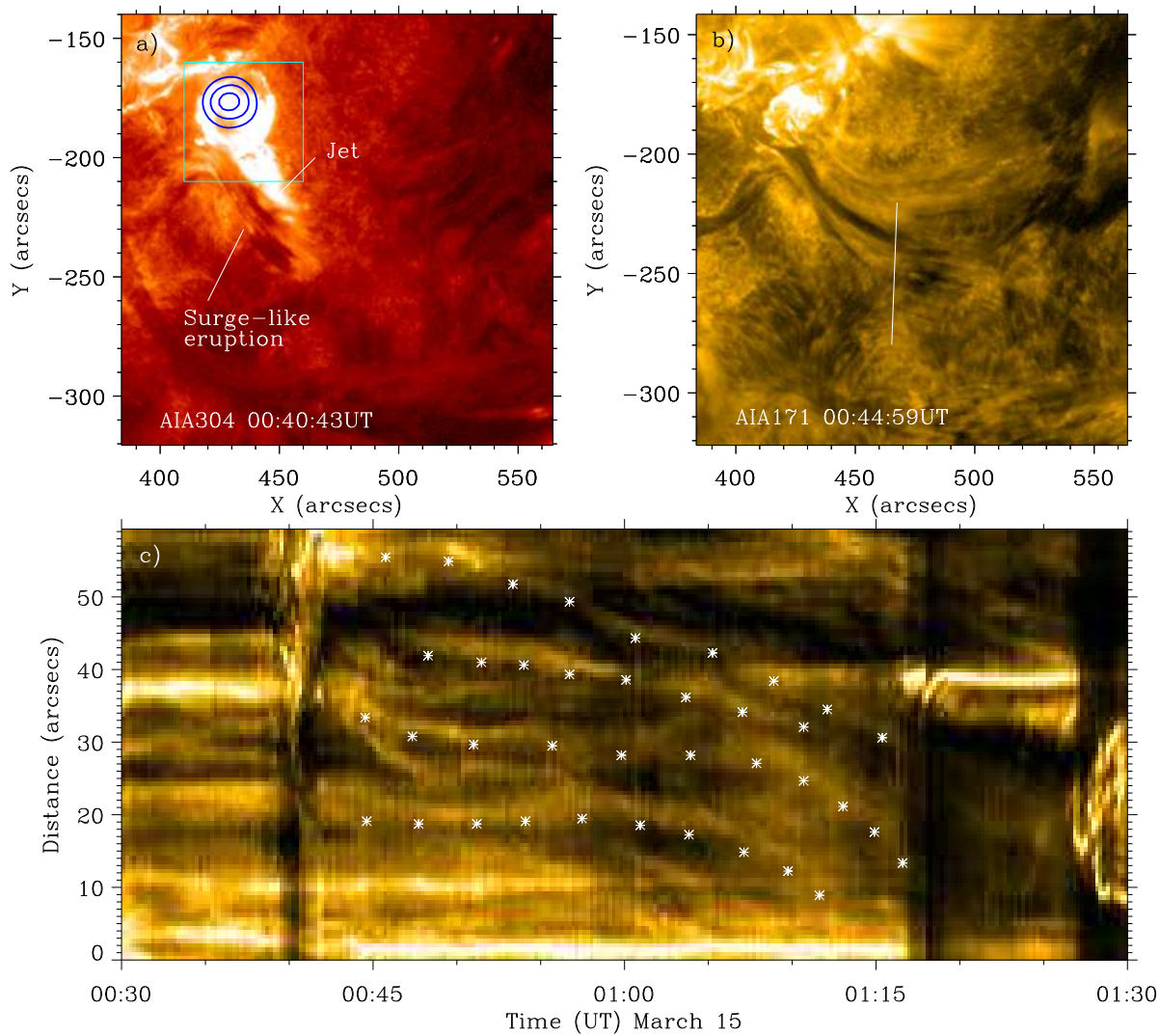}}
\caption{Characteristics of the F1 eruption. (a) Surge-like filament eruption associated with the C2.4 class flare and the X-ray jet, shown by the AIA 304 \AA~image overlaid with the RHESSI XR 6--12 keV map (blue contour) at 00:39 UT. The cyan box is the same as those in Figure 2. (b) Swing of the failed F1 eruption after the interaction with F2 in the AIA 171 \AA~image. The white line indicates the slice along the direction that is quasi-perpendicular to the erupting direction of F1. (c) Distance--time diagram created along the slice in (b). The asterisks outline the southward moving structure of the F1 eruption.}
\end{figure}
\clearpage
\begin{figure}
\centerline{\includegraphics[width=0.88\textwidth,clip=]{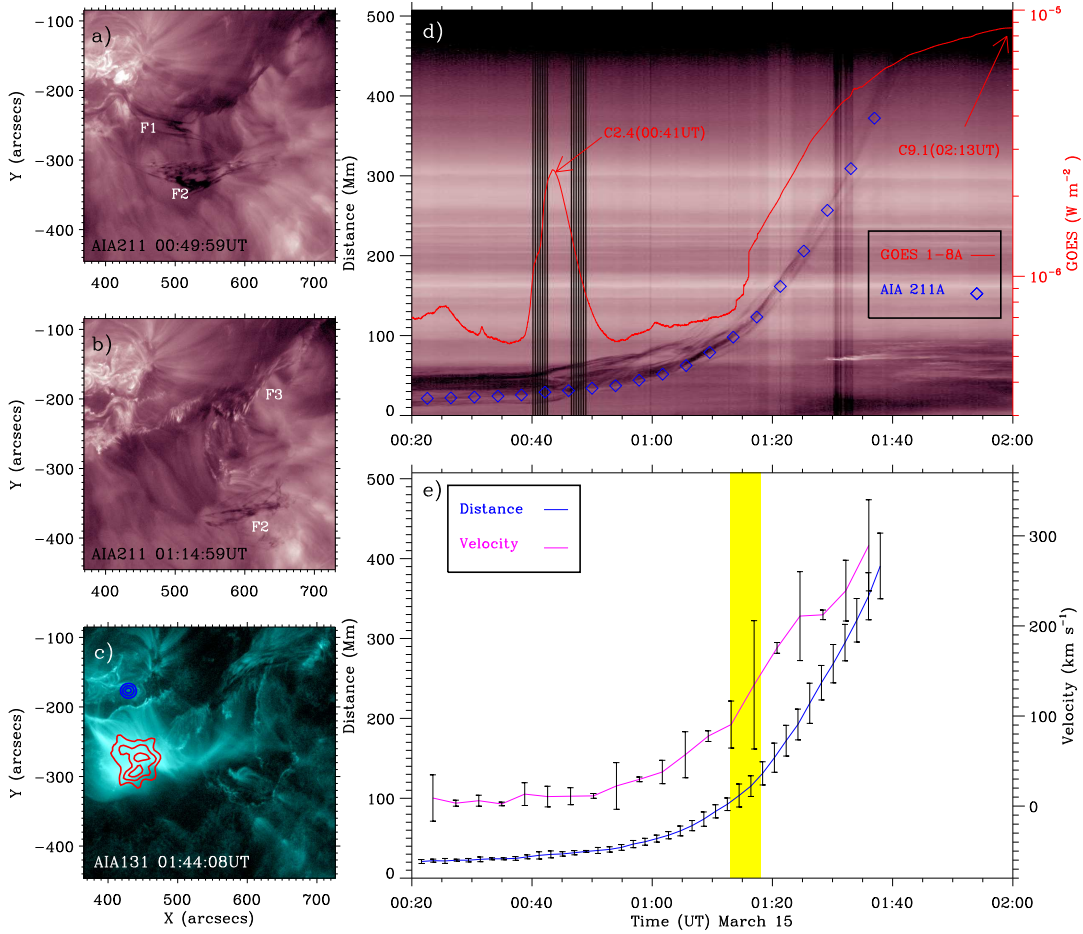}}
\caption{(a) Initiation of the F2 eruption by F1 in the AIA 211 \AA~image. (b) Acceleration of the erupting F2 and the induced F3 rise. (c) The flare loops below the erupting F2 in the AIA 131 \AA~image overlaid with XR 6--12 keV maps at 00:39 UT and 01:44 UT, which correspond to the sources of the C2.4 (blue contour) and C9.1 (red) flares, respectively. (d) Stack plot of the intensity along the slice as shown in Figure 1b. The {\it GOES} X-ray flux in the 1--8 \AA~channel is overplotted as the red curve. The peak times of the two flares are indicated. There were some gaps in the AIA data during the C2.4 flare. (e) Temporal evolution of the distances and velocities of F2. The yellow square shows the approximate onset time of the fast acceleration. The uncertainties of the distances and velocities are given as 3 standard deviations.}
\end{figure}
\clearpage
\begin{figure}
\centerline{\includegraphics[width=1\textwidth,clip=]{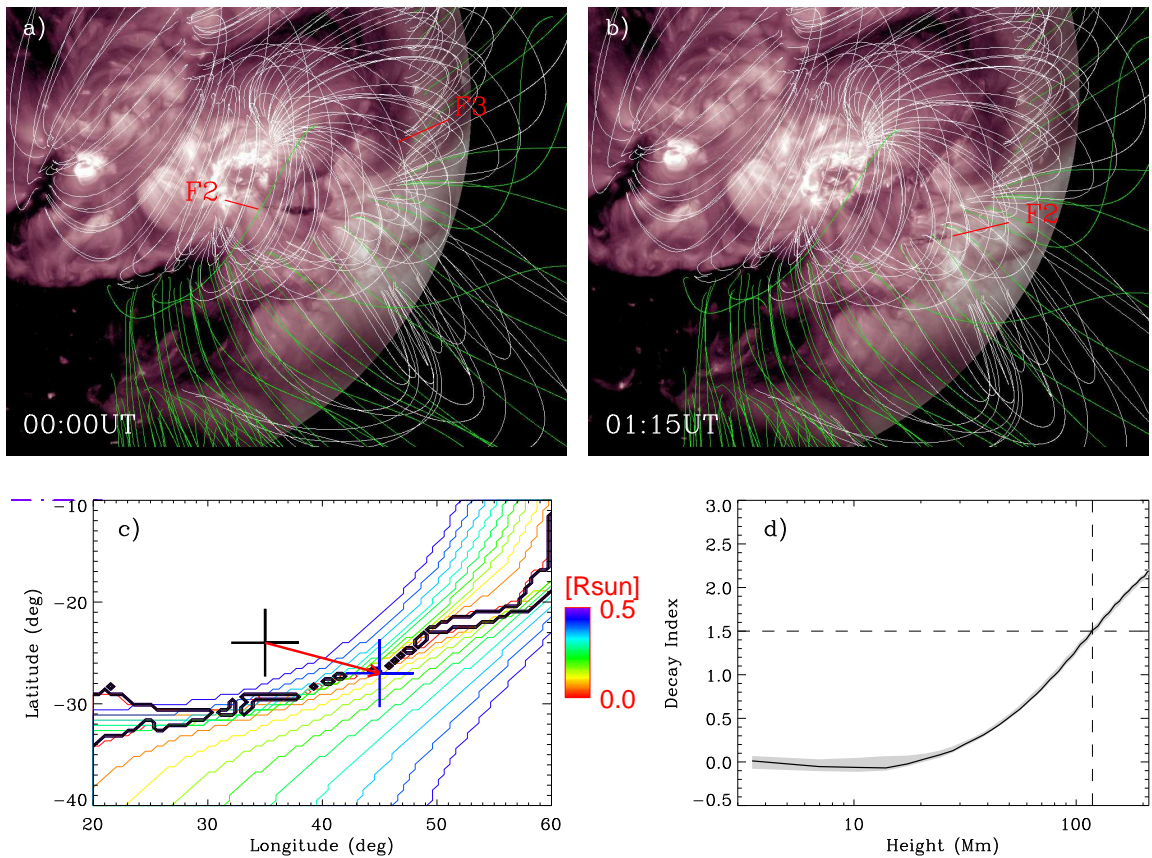}}
\caption{(a)-(b) PFSS reconstruction of the coronal magnetic field above the AR. The AIA 211 \AA~images at 00:00 UT and 01:15 UT are mapped on the solar surface as shown in panels (a) and (b), respectively. The white and green lines represent the closed and open magnetic fields, respectively. (c) Top view of the open magnetic field region with the projections of the colored contours. Different colors correspond to different heights. The black contours mark the footpoint belt of the open magnetic fields. The black and blue crosses in panel (c) correspond to the positions of F2 in panels (a) and (b), respectively. The red arrow indicates the propagation direction of F2. (d) Decay index of the magnetic field above F2. Black curve shows the median; shaded band indicates 1 $\sigma$ spread. Horizontal and vertical dashed lines indicate the critical value $n$ = 1.5 and critical height 118 Mm, respectively.}
\end{figure}
\clearpage




\end{document}